\documentclass[aps,prl,twocolumn]{revtex4}

\usepackage{graphicx}
\begin{document}

\title{Possibility of a Metallic Field--Effect Transistor}
\author{Slava V. Rotkin,
Karl Hess\\
Beckman Institute for Advanced Science and
Technology,\\
University of Illinois at Urbana--Champaign,\\
405 N.Mathews, Urbana,
IL 61801, USA;\\
Email: rotkin@uiuc.edu\\
Fax: (217) 244--4333}

\begin{abstract}

We develop theoretical arguments that demonstrate the possibility
of metallic field--effect transistors (METFET's) in
one--dimensional
 systems
 and particularly in
 armchair
carbon nanotubes. A very inhomogeneous electric field, such as
the field of a tunnelling tip, can penetrate the relatively
weakly screened nanotubes and open an energy gap. As a
consequence, an energy barrier forms that impedes electron flow
and thus permits transistor action. This type of metallic field
effect is advantageous because of the high conductance of the
metallic tubes in the ON--state.

\end{abstract}

%\twocolumn

%\begin{keywords}
%\end{keywords}

\maketitle

%_______________________________________
%\maketitle
%\twocolumn

%{\em Introduction. --}
Field--effect transistors in current use are semiconductor
devices. The scaling trend to nanometer dimensions calls for ever
higher doping and channel conductance of these devices
\cite{hessbook}. Ultimately one desires a conductance close to
that of a metal if one wants to scale devices to the smallest
possible size. However, a metallic conductance also prevents the
penetration of the electric field except for extremely short
distances; ordinarily too short to achieve device function.

We propose here an innovative approach to control electron
transport in {\em metallic one--dimensional} (1D) systems by use
of the {\em inhomogeneous} electric field of a low--dimensional
{\em highly localized} "modulating" gate (MG) such as nanometer
tips or metal nano-interconnects. Use of a highly localized gate
results in a strong enhancement of the electric field in a narrow
region. Thus any depletion of charge is easier to achieve since
it occurs only in extremely small volumes. In addition, the weak
screening of electric potentials in 1D channels further enhances
possible field effects.

The purpose of this letter is to show that metallic carbon
nanotubes may be suitable for use as metallic field--effect
transistors (METFET's). Key to this novel metallic field effect
is the opening of a band gap due to the breaking of the mirror
symmetry by the MG. Note, that a region of spatially non-uniform
electric field also represents a barrier that reflects %(scatters)
electrons, while the homogeneous field of an extended gate
changes predominantly the carrier density. These qualitative
factors are discussed below in a more quantitative way and their
potential for device applications is illustrated.

{\em Metallic field effect in Armchair NT's. --} IBM researchers
have enunciated the vision of combining metallic and
semiconducting single--wall nanotubes (M--SWNT's and S--SWNT's) in
circuits \cite{appenz-utah}. The M--SWNT's would serve as
interconnects and the S--SWNT's as active devices of the extremely
small size (typically, the SWNT radius is $\sim 0.7$~nm)
\cite{appenz2,appenz,avouris-r,avouris}. Here we discuss the
possibility to also use {\em metallic tubes} as transistors or
switching devices, which is appealing because of their virtually
ballistic conductance \cite{pseudos,lundstr,appenz2}.

\begin{figure}[!]
%\begin{figure}[h]
\centering
\includegraphics[width=3.in]{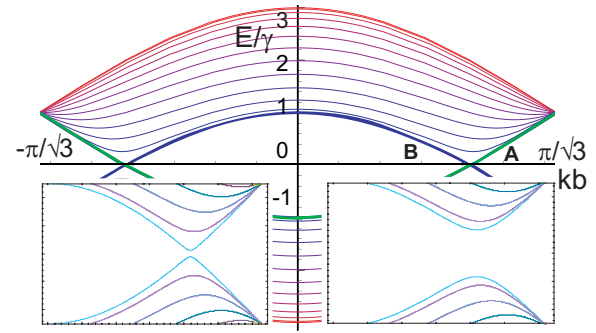}
\caption{ Band structure of armchair [10,10] M--SWNT. Left Inset
shows zoom view of the Fermi point with opening of a gap in the
M--SWNT. Right Inset shows how the gap grows linearly with applied
potential, cf. Eq.(1). } \label{fig:1}
\end{figure}

The band structure of an armchair SWNT, as shown in
Fig.\ref{fig:1}, has two subbands A and B (shown as bold green and
blue curves) crossing at the Fermi level. Absence of the gap is
ultimately related to the SWNT mirror symmetry \cite{damnjan},
which forbids any mixing of these subbands. We propose to break
this mirror symmetry by applying very non--uniform electric fields
by use of a MG. Two possible mechanisms for gap opening in
armchair SWNTs are: (1) by a direct mixing of electron and hole
states of the A and B subbands, which happens if the external
potential has an atomic scale corrugation, and (2) by the indirect
subband mixing in a higher order of the perturbation theory, which
essentially can be realized for any non--uniform potential. Both
mechanisms require (i) a relatively high electric field (order of
$10^7$ V/cm) and (ii) SWNT symmetry breaking.

Group theory proves that the direct matrix element is non--zero if
the Fourier transform of the potential $U_{q,\delta m}$ is a full
scalar with respect to the rotations of the symmetry group of the
nanotube \cite{damnjan}. This is fulfilled for $\delta m = s
\cdot n$, where $s=2,4,\dots$ is an positive even integer and $n$
is the integer appearing in the notation [n,n] for armchair
SWNT's \cite{newyan}. Then, the energy dispersion of two new
subbands $|A\pm B\rangle$ in vicinity of the Fermi point reads
as: $E_{\pm}= \pm\sqrt{E_A(q)^2+U_{q,\delta m}^2},$. Here
$E_A(q)=-E_B(q)=E_B(-q)$ is the dispersion in A/B subband at zero
gate voltage. At the Fermi point $E_A(0)=0$ and, thus, a gap
opens:
\begin{equation}
E_g^{(sn)}=2|U_{q,sn}|
\label{gap}
\end{equation}
This is a {\em direct mixing} of the subbands A and B. The bandgap
is linear in the applied gate potential (see Inset in Figure
\ref{fig:1}). A full calculation presenting a selfconsistent
solution of joint Schroedinger and Poisson equations following the
derivation of Refs. \cite{yanli,jetpl} will be published
elsewhere. Except for a natural condition that the field must not
cause an electric breakdown, there exists no upper limit for the
magnitude of the gap that opens in this case and we estimate the
possibility of a gap of several eV.

Thus direct mixing of the crossing subbands allows to open a large
gap. It does requires, however, a very high multipole moment of
the potential in order to break the symmetry (we must at least
have $\delta m = 2n$) and this is difficult to achieve because the
amplitude of any high multipole component will be typically much
smaller than the gate voltage, $U_{q,2n}<<V_g$. Nevertheless, it
may be possible to create such a potential by applying the gate
voltage to a chemically modified surface of the SWNT. This
approach is technologically complex.

\begin{figure}[!]
%\begin{figure}[h]
\centering
\includegraphics[width=3.in]{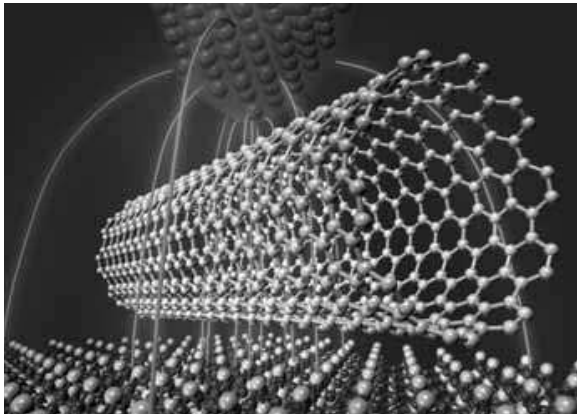}
\caption{Setup of an armchair SWNT METFET using the STM tip as a
modulation gate. } \label{fig:cover}
\end{figure}

\begin{figure}[!]
%\begin{figure}[h]
\centering
\includegraphics[width=3.5in]{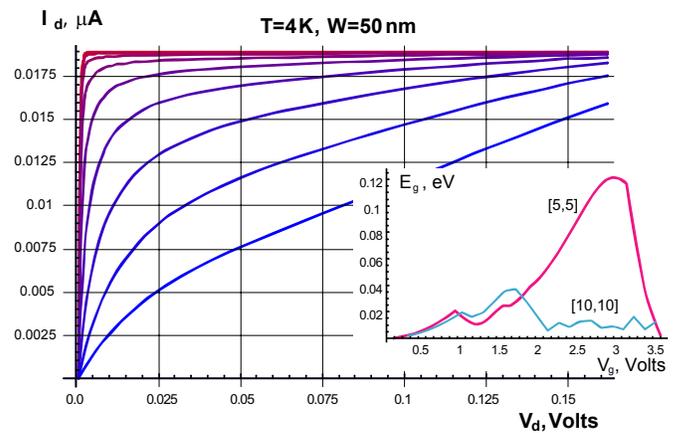}
\caption{ Current of a METFET versus the drain potential. The gate
width is 50 nm. $T=4$K. Curves from top to bottom correspond to
$V_g$ from 0 to 3 V (in color: from red to blue). Inset shows
corresponding bandgap opening for the same gate voltage range for
[5,5] and [10,10] SWNTs.} \label{fig:ivc}
\end{figure}

Fortunately, there exists a more straightforward possibility to
open a bandgap by virtually any kind of inhomogeneous electric
field that breaks the mirror symmetry. A bandgap opens also by
{\em indirect mixing} of the subbands. This bandgap opening is
smaller because it arises from higher order perturbation terms
(similar to a nonlinear Stark effect). In particular, any
potential that contains both a uniform field and a higher order
(quadrupole) component will lead to a gap which scales with the
third power of the gate voltage. In what follows we consider an
example of two metal tips at opposite sidewalls of the SWNT: one
at $+2 V_g$ and the other at $- V_g$ bias. The dependence of the
bandgap on $V_g$ is shown for [5,5] and [10,10] armchair SWNTs in
the Inset of Fig.\ref{fig:ivc}. The gap $E_g$ first increases with
$V_g$ and decreases beyond $\pi \gamma/(e n)$ where $\gamma\simeq
3$~eV is a hopping integral. Perturbation theory
\cite{yanli,newyan,nanotech} predicts a maximum gap that depends
only on the size of the tube and scales in a universal way for
nanotubes of arbitrary chirality:
\begin{equation}
\displaystyle E_g^{(c)}\sim
\frac{\hbar v_F b}{2 R^2} \propto
\frac{\gamma}{n^2}, \label{gapstark}
\end{equation}
This was confirmed by numerical calculations in
Refs.\cite{yanli,nanotech}. Here $v_F=3 \gamma b/2\hbar$ is the
Fermi velocity for a M--SWNT, and $b\simeq 0.14$~nm is the bond
length. Even though the maximum gap is small in this case and thus
has consequences for the conductance only at low temperatures, the
gap opening should be much easier to achieve experimentally e.g.
by use of the inhomogeneous electric field of a tunnelling tip
(Fig.\ref{fig:cover}). Also, ultra narrow leads \cite{rotkina},
fabricated closely to the nanotube channel, special
(electro--)chemical function groups at the tube sidewalls or
inside the tube \cite{zharov} may be used as a MG. We notice that
use of a dual gate (both local and backgate) may be beneficial for
1D METFET's because of the uniform backgate controls the Fermi
level (charge density) while the MG controls the conductivity of
the channel.

\begin{figure}[!]
%\begin{figure}[h]
\centering
\includegraphics[width=3.5in]{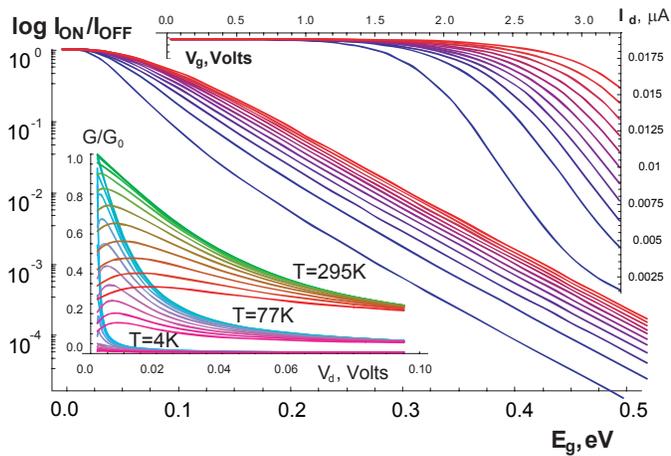}
\caption{The METFET current vs the opened bandgap. $W=50$ nm. The
red (upper) curve is at 295K, the blue (lowest) curve is at 4K.
The drain voltage dependence of the barrier transmittance is shown
in the Left Inset. The families of curves correspond different
temperatures: $T=$4K, 77K and 295K from bottom to top. Within each
family, the gap is increased from 0 to 0.12 eV from top to bottom
(in color: from blue--green to pink--red). Right Inset: METFET
current as a function of $V_g$ (at the bias $V_d$ from 0 to 0.15 V
from bottom to top, in color: from blue to red). }
\label{fig:transc}
\end{figure}

{\em Field effect modulation of metallic conductance. --} Any
opening of the gate induced semiconductor gap along the armchair
metallic tube will create a potential barrier for the electrons
and therefore modulate conduction. (Note that this modulation may
be enhanced by Coulomb blockade \cite{ftft}.) Inclusion of
tunnelling is important for the calculation of the current because
the effective mass of the electrons in SWNT's is very small ($\sim
0.06 m_o$, similar to the in--plane effective mass of graphite
\cite{ieeenano}). At non zero temperature thermionic emission must
be taken into account as well. Although the rate of tunnelling
through the classically forbidden (gated) region is high, a
significant metallic field effect can be achieved by increasing
the width of the gated region, or by operating the device at low
source--drain voltage, $V_d$, and low temperature, $T$. (The
ON/OFF current ratio is controlled by the parameter $E_g/V_d$.)
Because the transport in the armchair SWNT METFET is ballistic,
its conductance in the ON state is limited only by reflections at
the contact. We will use in what follows (mainly for normalization
purposes) a maximum conductance of $4G_o=2e^2/h$ \cite{ftn}.

For the OFF state, we estimate a semiclassical WKB transmission
coefficient, $T(E)$, by assuming a uniform gap of width $W$ in the
gated region. In order to extend the validity of our results to
non--zero temperatures we calculate the total transmission of the
channel by integrating the partial current, $\sim T(E)$,
multiplied by the difference of the Fermi distribution functions
of the left and right electrodes.

Figure \ref{fig:ivc} shows typical IV curves (IVC) for a METFET
using a [5,5] armchair M--SWNT of the diameter 0.7 nm at $T=$4K,
$W=50$~nm. The upper (red) IVC corresponds to a zero gate voltage
(no gap). The channel is fully open and the ON current is
determined by injection from a contact and thus by the quantum
conductance $4G_o$ \cite{ftn}. With increasing gate voltage one
observes a substantial decrease of the current due to the opening
of the gap and a depletion of electrons in this region (blue
curve). In the Right Inset of Fig.\ref{fig:transc}, the METFET
current is plotted vs. the $V_g$ for the dipole--quadrupole
potential (two tip geometry) as described above. At low drain bias
(blue curve) the METFET is switched OFF at $V_g \sim 3$~V.

The total transmission coefficient of the METFET, $T(E)=G(E)/G_o$,
assuming transparent contacts, is plotted in the Left Inset of
Fig.\ref{fig:transc}. Several families of curves are given for
different temperatures. In the upper family of curves (room
temperature) the suppression of transmission through the gated
region is clearly seen as a function of an increased gap: from 0
to 0.12 eV (from green to red, top to bottom). At lower
temperatures $T=$77K and 4K this effect is naturally sharper.

The current is an exponential function of the gap (and, hence, of
the gate voltage). Results are shown in Figure \ref{fig:transc}:
by opening a gap of about 0.5 eV (which would certainly require
sophisticated technology) one can decrease the current by five
orders of magnitude.

In conclusion, we have proposed a novel type of electronic
switching device based on carbon nanotubes: a one--dimensional
metallic field--effect transistor (1D METFET). We have described
armchair SWNT's in detail because of their special symmetry that
also gives rise to virtually ballistic transport. Our calculations
demonstrate, at least principle, the possibility to open a band
gap by application of inhomogeneous electric fields that are
created by special gates (modulation gates) that in the simplest
case resemble a tunnelling tip to break the SWNT symmetry.
Assuming ballistic transport, we have calculated IV curves for the
METFET's with gate widths of the order of 15--50 nm and found
significant modulation as well as reasonably large ON/OFF current
ratios. We finally note that one can expect excellent scaling
properties of the SWNT METFET because of the metallic conductivity
enhanced by ballistic transport.

%\begin{acknowledgments}
Authors are indebted to Mr.~B.~Grosser for help with image
preparation and Ms.~Y.~Li for help in numerical computations. This
work was supported by NSF Grant No. 9809520, by the Office of
Naval Research grant NO0014-98-1-0604 and the Army Research Office
grant DAAG55-09-1-0306. SVR acknowledges partial support of DoE
Grant No. DE-FG02-01ER45932, and NSF grant No. ECS--0210495.
%\end{acknowledgments}


\newpage
\begin{thebibliography}{99}

\bibitem{hessbook}
Advanced Theory Of Semiconductor Devices// Karl Hess. New York:
IEEE Press, 2000.


\bibitem{appenz-utah}
J. Appenzeller, Device Research Conference, Utah, June 26, 2003

\bibitem{appenz2}
J. Appenzeller, J. Knoch, R. Martel, V. Derycke, S.J. Wind, Ph.
Avouris,
%"Carbon nanotube electronics",
 IEEE Transactions on Nanotechnology,
{\bf 1} (4), 184, 2002.
%Volume: 1 Issue: 4 , Dec. 2002 Page(s): 184 -189

\bibitem{appenz}
J. Appenzeller, R. Martel, V. Derycke, M. Radosavljevic, S. Wind,
D. Neumayer, and Ph. Avouris,
%"Carbon nanotubes as potential building
%blocks for future nanoelectronics,"
Microelectronic Engineering, {\bf 64}, 391, 2002.
%pp. 391--397, 2002.

\bibitem{avouris-r}
Ph.G. Collins, Ph. Avouris,
%"Nanotubes for Electronics",
Sci.Am., {\bf 12}, 62%-69
, 2002.
Ph. Avouris,
%"Carbon Nanotube Electronics",
Chemical Physics, {\bf 281},
429%-445
, 2002.

\bibitem{avouris}
S. Heinze, J. Tersoff, R. Martel, V. Derycke,
J. Appenzeller, Ph. Avouris,
%"Carbon Nanotubes as Schottky Barrier Transistors"
Physical Review Letters {\bf 89}, 106801, 2002.


\bibitem{pseudos}
P.L. McEuen, M. Bockrath, D.H. Cobden, Y-G. Yoon, and S.G. Louie,
%"Disorder, Pseudospins, and Backscattering in Carbon Nanotubes"
Phys. Rev. Lett. {\bf 83}, 5098, 1999.
%5098–-5101 (1999)


\bibitem{lundstr}
 A. Javey, H. Kim, M. Brink, Q. Wang, A. Ural, J. Guo, P.
McIntyre, P. McEuen, M. Lundstrom, H. Dai,
%"High-[kappa] dielectrics for advanced
%carbon-nanotube transistors and logic gates"
 Nature Materials {\bf 1},
241, 2002.
%241--246 (01 Dec 2002)




\bibitem{damnjan}
T. Vukovic, I. Milosevic, M. Damnjanovic,
%"Carbon nanotubes band assignation, topology, Bloch states,
%and selection rules",
Physical Review {\bf B 65}(04), 5418, 2002.


\bibitem{newyan}
Y.Li, S.V.Rotkin, U.Ravaioli, and K.Hess,
%"Mirror symmetry breaking and band gap in
%armchair metallic carbon nanotubes",
unpublished.



\bibitem{yanli}
Y.Li, S.V.Rotkin and U.Ravaioli,
%"Electronic response and bandstructure modulation
%of carbon nanotubes in a transverse electrical field",
Nano Letters, {\bf 3}(2), %183--187, 2003.
183, 2003.

\bibitem{jetpl}
K.A. Bulashevich, S.V. Rotkin,
JETP Letters {\bf 75}%(4)
, 205 %--209
, 2002.


\bibitem{nanotech}
S.V. Rotkin, K. Hess, Proc. of Nanotech 2004, Boston, March 7-11,
2004 (in press).


\bibitem{rotkina}
J.-F. Lin, J.P. Bird, L. Rotkina, P.A. Bennett,
%"Classical and quantum transport in
%focused-ion-beam-deposited Pt nanointerconnects",
 Appl. Phys. Lett. {\bf 82} (5), 802, 2003.
%Applied Physics Letters -- February 3, 2003 --
%Volume 82, Issue 5, pp. 802-804




\bibitem{zharov}
S.V. Rotkin, I.Zharov,
%"Nanotube Light-Controlled Electronic Switch",
%Published by: World Scientific, Singapore.
Int. Journal of Nanoscience {\bf 1} (3/4), %347--356, 2002.
347, 2002.

\bibitem{ftft}
At the points of metal--semiconductor junction (MSJ) a charge
accumulation may happen. This MSJ boundary becomes a side of a 1D
capacitor with a capacitance $\sim W/2\ln (W/R)$. Neglecting the
logarithmic term, a classical charging energy of this capacitor is
$e^2/2W$, which is large for a narrow gate. However, in this work
we did not consider a Coulomb blockade because an effective gap is
exponentially renormalized at a high conductance of the tunnel
junction \cite{matveev}.
%The specifics of nanotube
%MSJ is that for SWNT band parameters
%(very light effective mass)
%the tunnelling
%%length is large (from sub--nm to
%%tens of nm) and the tunnelling
%rate is high. For
Large
coupling between sides of the MSJ
and quantum fluctuations of the
charge wash out the correlation and destroy the Coulomb
blockade.

\bibitem{matveev}
K.A. Matveev,
%"Coulomb blockade at almost perfect transmission",
 Phys. Rev. {\bf B 51},
1743, 1995.
%1743-–1751 (1995)


\bibitem{ieeenano}
The effective mass near the band edge $m^*=\hbar^2(\partial^2
E/\partial k^2)^{-1}=2\hbar^2 /(9b R \gamma)$ is about 0.06 of the
free electron mass for the SWNT of the radius $R\simeq 0.7$~nm.
%Here $\gamma\simeq 2.7$~eV
%is a hopping integral and
%$b\simeq 0.14$~nm is a SWNT
%bond length.


\bibitem{ftn}
For a circuit with macroscopic leads to the M--SWNT channel the
total conductance will be about $4G_o=2e^2/h$, 4 times of the
conductance quantum (for 2 spin and 2 space channels). This gives
a minimum resistance
 of the SWNT device
$\sim 6.5 k\Omega$. The lower resistance can be expected in the
case of entirely nanotube circuit \cite{appenz-utah}. The quantum
contact resistance will not limit anymore the ON current in this
case. That device can fully exploit all advantages of the METFET.








\end{thebibliography}
\end{document}